\documentclass[aps,prl,twocolumn,superscriptaddress,longbibliography]{revtex4-1}
\usepackage{hyperref} 
\usepackage{graphicx}
\usepackage{bm}
\usepackage{amsmath}
\usepackage{amssymb}
\usepackage{braket}
\usepackage[colorinlistoftodos]{todonotes}
\usepackage{xcolor}

\DeclareGraphicsExtensions{.png,.jpg,.eps}

\begin{document}
\author{Wei Luo}
\affiliation{National Laboratory of Solid State Microstructures, School of Physics,
and Collaborative Innovation Center of Advanced Microstructures, Nanjing University, Nanjing 210093, China}

\author{Wei Chen}
\email{Corresponding author: pchenweis@gmail.com}
\affiliation{National Laboratory of Solid State Microstructures, School of Physics,
and Collaborative Innovation Center of Advanced Microstructures, Nanjing University, Nanjing 210093, China}

\author{D. Y. Xing}
\affiliation{National Laboratory of Solid State Microstructures, School of Physics,
and Collaborative Innovation Center of Advanced Microstructures, Nanjing University, Nanjing 210093, China}

\title{Anomalous Electromagnetic Induction Engendered by Singular Gauge Transformation}
\begin{abstract}
The Berry curvature, resembling the magnetic field in
reciprocal space, offers a captivating avenue for exploring unique
electromagnetic phenomena devoid of real-space analogs.
Here, we investigate the emergent electromagnetic induction by
solenoidal Berry curvature with its field lines forming loops,
links, and knots. In stark contrast to Faraday's law, which dictates that alternating
magnetic fields yield alternating electric fields with a net zero average, the
alternating Berry curvature can engender
directional electromagnetic induction. Such an effect is attributed to
the presence of singularities in the Berry curvature,
accompanied by a $2\pi$ jump in the Berry flux. Notably, this jump does not trigger a diamagnetic impulse,
due to the gauge invariance of the Berry phase modulo $2\pi$. Consequently,
the induced electric field maintains finite values under time averaging, manifesting
itself as a directional pumping current. Our research sheds light on an anomalous
electromagnetic induction effect directly arising from the singular
gauge transformation, thereby expanding our comprehension of exotic
electromagnetic phenomena.
\end{abstract}

\date{\today}

\maketitle

The Berry phase effect has become an important research field
in modern physics~\cite{berry84,xiaod10rmp}, akin to the generalized
Aharonov-Bohm phase~\cite{AB59pr} in parametric spaces.
In solid-state physics, the Berry phase of the Bloch wave function
accrues along closed trajectories in reciprocal space.
Interestingly, physical properties such as the semiclassical dynamics of the Bloch electrons
exhibit captivating duality between real and reciprocal spaces~\cite{Chaudhary18prb,girvin2019modern},
thereby offering a tangible perspective for understanding the topological
properties in reciprocal space. Specifically, one can envision the Berry curvature
as a local magnetic field in reciprocal space, imparting
a wealth of intriguing characteristics to Bloch electrons~\cite{xiaod10rmp}.
Drawing inspiration from this analogy, novel effects traditionally confined to
real space have recently found their implementation in reciprocal space.
Noteworthy examples include the Aharanov-Bohm effect~\cite{flaschner16sci,duca15sci}
and Hall drift~\cite{jotzu16nat}.

In addition to the physical effects that have counterparts in real space,
there are also electromagnetic phenomena that are exclusive to reciprocal space.
A prime example is the existence of magnetic monopoles in the
reciprocal space of Weyl semimetals, serving as sources or drains for
the Berry curvature~\cite{wan11prb,armitage18rmp}, while
the corresponding object in real space remains elusive despite extensive pursuit.
Mathematically, incorporating the effects of magnetic monopoles into the modified
Maxwell equations is straightforward~\cite{jackson1999classical}.
As a direct physical consequence, these Berry curvature monopoles can lead to
anomalous electromagnetic induction (EMI) and resultant charge-pumping
effects in Weyl semimetals~\cite{PhysRevLett.117.216601,PhysRevB.95.245211,ma2017direct}.
In this context, an interesting question that arises is whether there are exotic emergent
electromagnetic phenomena in reciprocal space that extend beyond the framework of Maxwell's theory.


\begin{figure}[t!]
\begin{center}
\includegraphics[width=1\columnwidth]{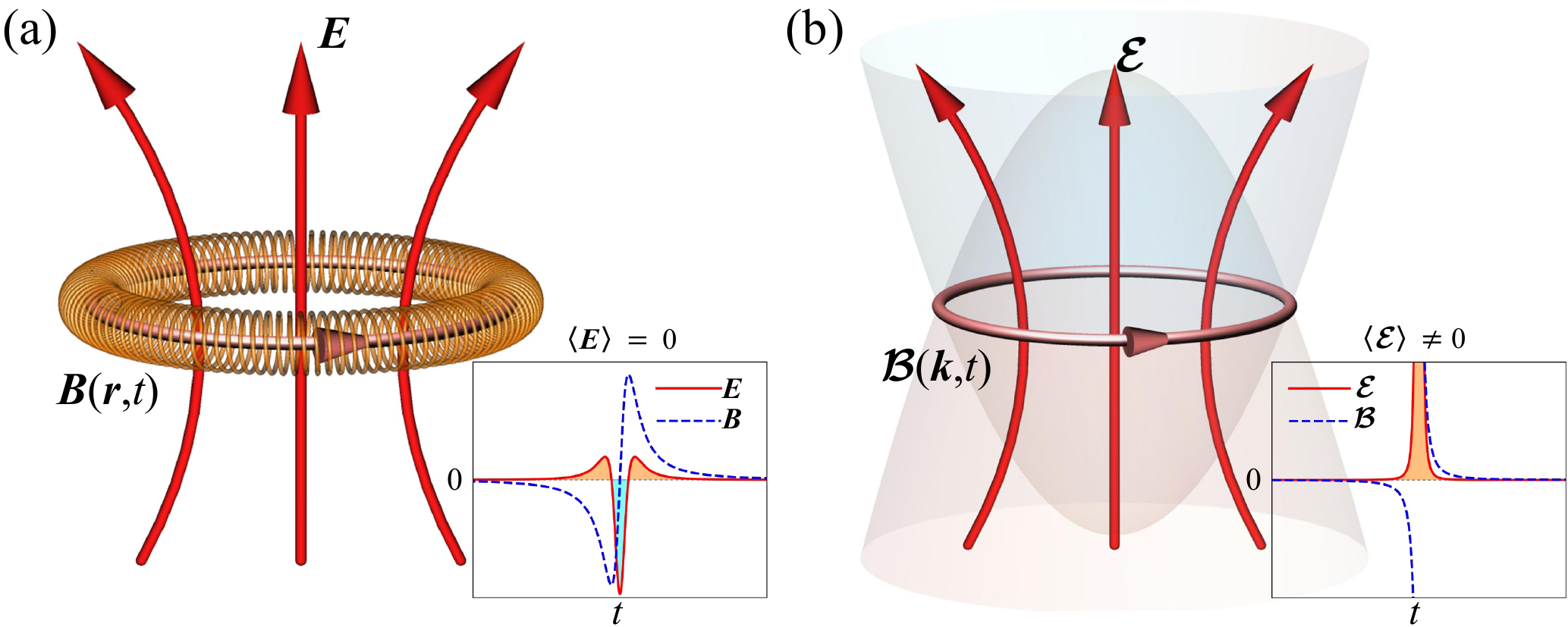}\\[0pt]
\caption{(a) Schematic illustration of conventional electromagnetic induction:
an alternating solenoidal magnetic field $\bm{B}(\bm{r},t)$ induces an
alternating electric field $\bm{E}(\bm{r},t)$ with a zero time-averaged value.
An abrupt change in $\bm{B}(\bm{r},t)$ results in a strong diamagnetic impulse, as shown in the inset.
(b) Anomalous electromagnetic induction: the induction of an emergent electric field
$\bm{\mathcal{E}}(\bm{k},t)$ by solenoidal Berry curvature $\bm{\mathcal{B}}(\bm{k},t)$
with finite distributions around band crossing regions. Importantly, an abrupt change
in the Berry flux by $2\pi$ does not lead to a diamagnetic electric impulse, resulting
in non-zero average values of $\bm{\mathcal{E}}(\bm{k},t)$.}
\label{Fig.1}
\end{center}
\end{figure}

In this work, we affirmatively answer this question by incorporating
singularities in the theory of electrodynamics.
Specifically, we investigate the emergent
electromagnetic induction (EMI) generated by the time-varying solenoidal
(divergence-free) Berry curvature $\bm{\mathcal{B}}(\bm{k},t)$, as illustrated in Fig.~\ref{Fig.1}(b).
It can be proved that the emergent electric field $\bm{\mathcal{E}}(\bm{k},t)$
and $\bm{\mathcal{B}}(\bm{k},t)$ satisfy the conventional Faraday's law of EMI, expressed as
$\nabla_{\bm{k}}\times \bm{\mathcal{E}}=-\partial\bm{\mathcal{B}}/\partial t$.
One might initially expect that a periodically variation of the Berry curvature
would induce an alternating $\bm{\mathcal{E}}$ field with a zero time-averaged value,
as its real-space counterpart shown in Fig.~\ref{Fig.1}(a).
However, this conclusion holds true only when the Berry curvature $\bm{\mathcal{B}}$ varies
continuously with time, a characteristic inherent in a real magnetic field
$\bm{B}$ but not in $\bm{\mathcal{B}}$. Notably, the
Berry curvature $\bm{\mathcal{B}}(\bm{k},t)$ in reciprocal space can accommodate singularities and
abrupt changes in its flux by multiples of $2\pi$.
Given that the Berry phase is gauge-invariant modulo 2$\pi$,
abrupt shifts in the Berry flux from $-\pi$ to $\pi$ do not induce
diamagnetic electric impulses, as seen in the comparison between
the insets in Figs.~\ref{Fig.1}(a) and ~\ref{Fig.1}(b).
Consequently, due to the absence of such counteractive impulses, the
$\bm{\mathcal{E}}$ field possesses a finite value under time averaging, which can
be probed through the charge pumping effect.
Based on this argument, it becomes apparent that this directional
EMI induced by alternating solenoidal Berry curvature
is a direct manifestation of the singular gauge
transformation (SGT).
Furthermore, we extend our investigation to explore the emergent
EMI effect induced by other solenoidal Berry curvature with its field lines forming
links and knots, thereby enriching the resultant topological pumping effect.
Our theory unveils an anomalous electromagnetic phenomenon in reciprocal space, devoid of
any real-space counterparts.

\begin{figure}[t!]
    \begin{center}
    \includegraphics[width=1\columnwidth]{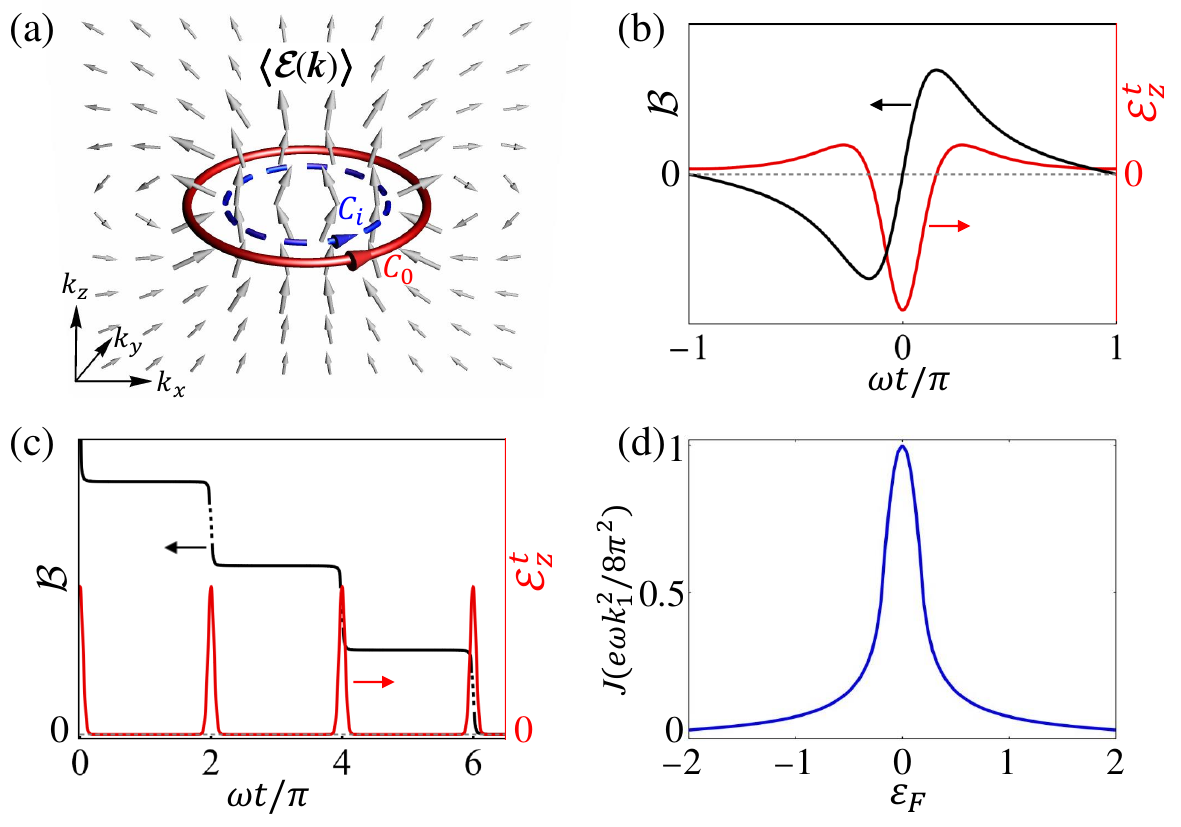}\\[0pt]
    \caption{
(a) Illustration of the singular loop $C_0$ (red circle) and a
regular loop $C_i$ (blue circle), with arrows indicating the direction
of the Berry curvature. The time-averaged distribution
of the $\bm{\mathcal{E}}(\bm{k})$ field is represented by the
gray vectors. Only the distribution in the $k_x$-$k_z$ plane is shown for clarity,
and its entire distribution possesses rotational symmetry around the $k_z$ axis.
(b) Time dependence of the Berry curvature on the loop with radius
$k_{\rho}=0.85k_1$ lying in the $k_z=0$ plane and the $z$ component
of the induced transverse electric field $\mathcal{E}_z^t$ at the origin.
(c) Time dependence of the Berry curvature on the red loop $C_0$ after
singular gauge transformation and the induced $\mathcal{E}_z^t$ at the origin.
(d) Pumping current density as a function of the Fermi energy $\varepsilon_F$.
The parameters are set to $g_1=2$, $g_2=2$, $k_0=1.3$, $\lambda=1$, and $\eta=1$.}
    \label{Fig.2}
    \end{center}
    \end{figure}

An elementary example of solenoidal Berry curvature can be
realized through a linear band crossing, as shown in Fig.~\ref{Fig.1}(b),
accompanied by a mass term. This system is described by the Hamiltonian given as
\begin{equation}\label{Ham NLSM}
H(\bm{k},t)=\lambda k_z \sigma_x+ g_y(t){\sigma _y} +\left[\eta(k_0^2-k^2)+g_z(t)\right]\sigma_z,
\end{equation}
where $k=\sqrt{k_x^2+k_y^2+k_z^2}$ is the magnitude of the momentum,
the Pauli matrices $\sigma_{x,y,z}$
act on the (pseudo-)spin, $\lambda$ and $\eta$ are model parameters.
The two driving terms $g_y(t)={g_1}\sin (\omega t)$ and $g_z(t)={g_2}\cos (\omega t)$
introduce periodic modulations to the Berry curvature, leading to
the EMI effect.
The instantaneous eigenvalues of the Hamiltonian are
$\varepsilon_{\alpha}(\bm{k},t)=\alpha\sqrt{\lambda^{2}k_{z}^{2}+g_{y}(t)^{2}
+[\eta(k_{0}^{2}-k^{2})+g_{z}(t)]^{2}}$, with corresponding eigenstates represented as
$|\chi_{\alpha}(\bm{k},t)\rangle$, where $\alpha=\pm$ denotes
the conduction and valence bands. We assume $g_2>\eta k_0^2$ to ensure that
the two bands periodically intersect at $t=2n\pi/\omega$ along the circular nodal loop,
denoted by $C_0$, having a radius $k_1=k_0[1+g_2/(\eta {k_0^2})]^{1/2}$
in the $k_z=0$ plane, as depicted in Fig.~\ref{Fig.2}(a).
The time-dependent Berry curvature for the valence band is calculated by
$\bm{\mathcal{B}}(\bm{k},t)=\nabla_{\bm{k}}\times\bm{\mathcal{A}}$
with $\bm{\mathcal{A}}=i\langle\chi_-|\nabla_{\bm{k}}|\chi_-\rangle$ the Berry
connection, which yields
\begin{equation}\label{Berry curvature NLSM}
\bm{\mathcal{B}}(\bm{k},t)=\lambda \eta (k_x^2+k_y^2)^{1/2}g_{1}\sin(\omega t)
\hat{\bm{e}}_{\theta}/\varepsilon_+^{3}.
\end{equation}
Note that $\bm{\mathcal{B}}$ possesses only the azimuth component with its
direction indicated by the unit vector $\hat{\bm{e}}_{\theta}$. Moreover, it
converges into divergent $\pm\pi$ Berry fluxes along the nodal loop at $t=2n\pi/\omega\pm0^+$ for any integer $n$.

We solve the time evolution of the wave function
to investigate the physical effect induced by the variation of the Berry curvature.
Without loss of generality, we consider an initial state
$|\chi_{-}(\bm{k},0)\rangle$ in the valence band,
which evolves into $|\psi_{-}(t)\rangle=|\chi_{-}(t)\rangle-
i\hbar\frac{\langle\chi_{+}|\partial_{t}\chi_{+}\rangle}
{\varepsilon_{-}-\varepsilon_{+}}|\chi_{+}(t)\rangle$ to first-order approximation.
This evolution arises from the gradual changes
in both the momentum $\bm{k}$ and the driving terms $g_{y,z}(t)$.
The time derivative in the second term is explicitly expressed as $|\partial_t\chi_\alpha\rangle=\dot{\bm{k}}\cdot\nabla_{\bm{k}}|\chi_\alpha\rangle
+\sum_{j=y,z}\dot{g}_{j}\partial_{g_{j}}|\chi_\alpha\rangle$.
The average value of the velocity $\dot{\bm{r}}(\bm{k},t)=\nabla_{\bm{k}}H(\bm{k},t)$
with respect to the state $|\psi_-(t)\rangle$ is obtained as~\cite{Chaudhary18prb}
\begin{equation}\label{velocity}
    \dot{\bm{r}}=\nabla_{\bm{k}}\varepsilon_-+\dot{\bm{k}}\times\bm{\mathcal{B}}
    +\bm{\mathcal{E}},
\end{equation}
where the last term is expressed as
\begin{equation}\label{electric field 1}
    \bm{\mathcal{E}}=i\sum_{j=y,z}\dot{g}_{j}\left(\langle\nabla_{\bm{k}}\chi_-|\partial_{g_j}
    \chi_-\rangle-\langle\partial_{g_j}\chi_-|\nabla_{\bm{k}}\chi_-\rangle\right).
\end{equation}
It can be decomposed into transverse and longitudinal components as
$\bm{\mathcal{E}}=\bm{\mathcal{E}}^{\text{t}}+\bm{\mathcal{E}}^{\text{l}}$.
Eq.~\eqref{velocity} takes on a similar form to
its dual equation $\dot{\bm{k}}=\dot{\bm{r}}\times\bm{B}+\bm{E}$, which
describes the momentum change of a charged particle (with unity charge)
due to the electromagnetic fields $\bm{B}$ and $\bm{E}$. By drawing this analogy, the Berry
curvature $\bm{\mathcal{B}}$ resembles the magnetic field, while
$\bm{\mathcal{E}}$ can be regarded as the emergent electric field in reciprocal space~\cite{PhysRevLett.117.216601}.
Interestingly, these fictitious fields in reciprocal space also adhere to
Faraday's law, as can be confirmed by taking the curl on both sides of
Eq.~\eqref{electric field 1}.
Substituting $|\chi_-\rangle$ into Eq.~\eqref{electric field 1}, we obtain the
three components of the $\bm{\mathcal{E}}$ field
\begin{equation}\label{ef}
\begin{split}
&\mathcal{E}_{x(y)}=\lambda g_{1}\omega \eta k_{x(y)}k_{z}\cos(\omega t)/\varepsilon_+^{3},\\
&\mathcal{E}_{z}=\lambda g_{1}\omega\left[g_{2}+\eta(k_{0}^{2}-k^{2}+2k_{z}^{2})
\cos(\omega t)\right]/(2\varepsilon_+^{3}).
\end{split}
\end{equation}
In Fig.~\ref{Fig.2}(a), we illustrate the time-averaged distribution of $\bm{\mathcal{E}}(\bm{k})$
in reciprocal space, resembling the electric field induced by the varying solenoidal magnetic field.

Here, the key result is that even though the Hamiltonian and
the resultant Berry curvature $\bm{\mathcal{B}}$ exhibit periodic variations
over time, the induced $\bm{\mathcal{E}}$ field and its transverse
component $\bm{\mathcal{E}}^{\text{t}}$ maintain non-zero average values.
This outcome stands in contrast to Faraday's law, which dictates that an
alternating magnetic field generates an alternating electric field with an
average value of zero. This underscores a clear differentiation between
electromagnetic phenomena in reciprocal and real spaces.
A crucial point is that the Berry curvature $\bm{\mathcal{B}}$
can harbor singularities. It is the SGT, characterized by an abrupt
change in the Berry flux carried by the singular Berry curvature
($\Phi: -\pi\rightarrow\pi$), that leads to a finite average value
of $\bm{\mathcal{E}}$, a phenomenon absent in its real-space counterpart.

To illustrate this point, let us begin by focusing on the transverse field $\bm{\mathcal{E}}^{\text{t}}$,
which is solely induced by $\bm{\mathcal{B}}$.
Due to the rotational invariance of the system
about the $z$ axis, the distribution of $\bm{\mathcal{B}}$ can be broken down into
various circular loops centered on the $k_z$ axis, as depicted in Fig.~\ref{Fig.2}(a).
The variations of $\bm{\mathcal{B}}$ in these loops
induce $\bm{\mathcal{E}}^{\text{t}}$.
For most of these circular loops, denoted by $C_i$, where no
singularity in $\bm{\mathcal{B}}$ occurs during each driving cycle, the average value of
$\bm{\mathcal{E}}^{\text{t}}$ is zero, following the Biot-Savart law
$\langle\bm{\mathcal{E}}^{\text{t}}(\bm{k},{C_i})\rangle=-\frac{\Delta \mathcal{B} \delta A}{T}\oint_{C_i}\frac{d\bm{k}'\times(\bm{k}-\bm{k'})}
{|\bm{k}-\bm{k'}|^{3}}=0$. Here, $d\bm{k}'$ is the vector element along $C_i$, $T=2\pi/\omega$ is the
driving period, and $\delta A$ denotes the cross-sectional area of the tiny circular tube enclosing $C_i$.
Since the magnitude of the
Berry curvature $\mathcal{B}$ is a regular function of time, its variation
within one period is zero, i.e., $\Delta\mathcal{B}=0$. This results in $\langle\bm{\mathcal{E}}^{\text{t}}(\bm{k},{C_i})\rangle=0$,
which can be confirmed by Fig.~\ref{Fig.2}(b).

In contrast to the $C_i$ loops, $\bm{\mathcal{B}}$ exhibits periodic
divergence along the $C_0$ loop and undergoes an abrupt change in its Berry flux by $2\pi$.
Given that the Berry phase is gauge-invariant modulo $2\pi$, the $\pm\pi$ Berry
fluxes are essentially equivalent. Leveraging this property, we can re-express
the periodically varying Berry curvature along the $C_0$ loop as follows
\begin{equation}
    \tilde{\bm{\mathcal{B}}}_0(\bm{k},t)\equiv \bm{\mathcal{B}}_0(\bm{k},t)-\left[\sum_{n=1}^\infty \Theta(t-nT)\right]\delta\bm{\mathcal{B}}(\bm{k})\hat{\bm{e}}_{\theta},
\end{equation}
where $\delta\bm{\mathcal{B}}(\bm{k})=2\pi\delta(k_z)\delta\left[k_1-(k_x^2+k_y^2)^{1/2}\right]$ captures
the change in Berry flux and $\Theta$ is the unit step function. This SGT effectively converts the periodic function
into a continuous, monotonically decreasing function characterized by a series of steps,
as shown in Fig.~\ref{Fig.2}(c). Each
jump of $\tilde{\bm{\mathcal{B}}}_0$ induces an identical pulse
in $\bm{\mathcal{E}}^{\text{t}}$. Therefore, it is the SGT
that enables directional EMI by the alternating Berry curvature.

The directional EMI induced by the SGT can be manifested as
charge pumping driven by the emergent field $\bm{\mathcal{E}}$.
We first consider a fully occupied valence band.
The average current density, denoted as $\bm{J}$, can be
obtained by integrating the velocity $\dot{\bm{r}}(\bm{k},t)$
across the entire band, which simplifies to
\begin{equation}\label{current}
    \bm{J}=\frac{e}{T}\int_0^T dt\int\frac{d^3\bm{k}}{(2\pi)^{3}}\bm{\mathcal{E}}(\bm{k},t).
\end{equation}
It is worth noting that the contributions from the first two terms in
Eq.~\eqref{velocity} both vanish after integration.
This implies that the current density is directly proportional to the
average $\bm{\mathcal{E}}$ field in reciprocal space.
Inserting Eq.~\eqref{ef} into Eq.\ (\ref{current}) yields
\begin{equation}\label{current 2}
\bm{J}=\frac{e\omega}{8\pi^{2}}k_{1}^{2}{\hat{\bm {z}}}.
\end{equation}
Here, the current flows in the $z$ direction since only $\mathcal{E}_z$ has a nonzero average value.
This result can be understood straightforwardly by breaking down the 3D system into multiple 1D
channels in reciprocal space, with each channel labeled by
$(k_x, k_y)$ and oriented in the $z$ direction. Eq.~\eqref{current 2} implies
that each 1D channel within the nodal loop with $k_\rho=\sqrt{k_x^2+k_y^2}<k_1$ makes an equal contribution to the current.
This result aligns with the characteristics of $\langle\bm{\mathcal{E}}(\bm{k})\rangle$ presented in Fig.~\ref{Fig.2}(a):
$\mathcal{E}_z$ maintains consistently positive as $k_z$ varies for $k_{\rho}<k_{1}$,
whereas it changes sign as $k_z$ varies for $k_{\rho}>k_{1}$, ultimately resulting
in an average value of zero. To be precise,
when we perform partial integration of $\frac{1}{2\pi}\int dk_zdt\mathcal{E}_z$
in Eq.~\eqref{current}, we obtain a Chern number of 1 for
$k_{\rho}<k_{1}$ and a Chern number of 0 for $k_{\rho}>k_{1}$.

In more general scenarios, the electron bands are only partially occupied. For instance, we consider
the valence band initially ($t=0$) filled up to a certain negative energy level $\varepsilon_F<0$.
Additionally, we focus on the adiabatic regime where the electron occupation remains constant during the
time evolution. This requirement implies that the pumping frequency $\omega$ is much higher than
the inverse relaxation time $1/\tau$. Under these conditions, the pumping current can still
be calculated using Eq.~\eqref{current}, with the integration now performed over the
occupied states rather than the entire band.
Similarly, for a positive Fermi energy $\varepsilon_F>0$,
the contribution from the conduction band should also be taken into account.
However, in both of these partially occupied scenarios,
no quantized expression like Eq.~\eqref{current 2} is available and so
we present the numerical results in Fig.~\ref{Fig.2}(d).
It shows that the pumping current decreases as the Fermi distribution deviates from
the fully occupied state of the valence band. The symmetrical structure around
zero energy signifies that both the energy levels and the emergent
$\bm{\mathcal{E}}$ field have opposite signs for the two bands.

\begin{figure}[t!]
\begin{center}
\includegraphics[width=1\columnwidth]{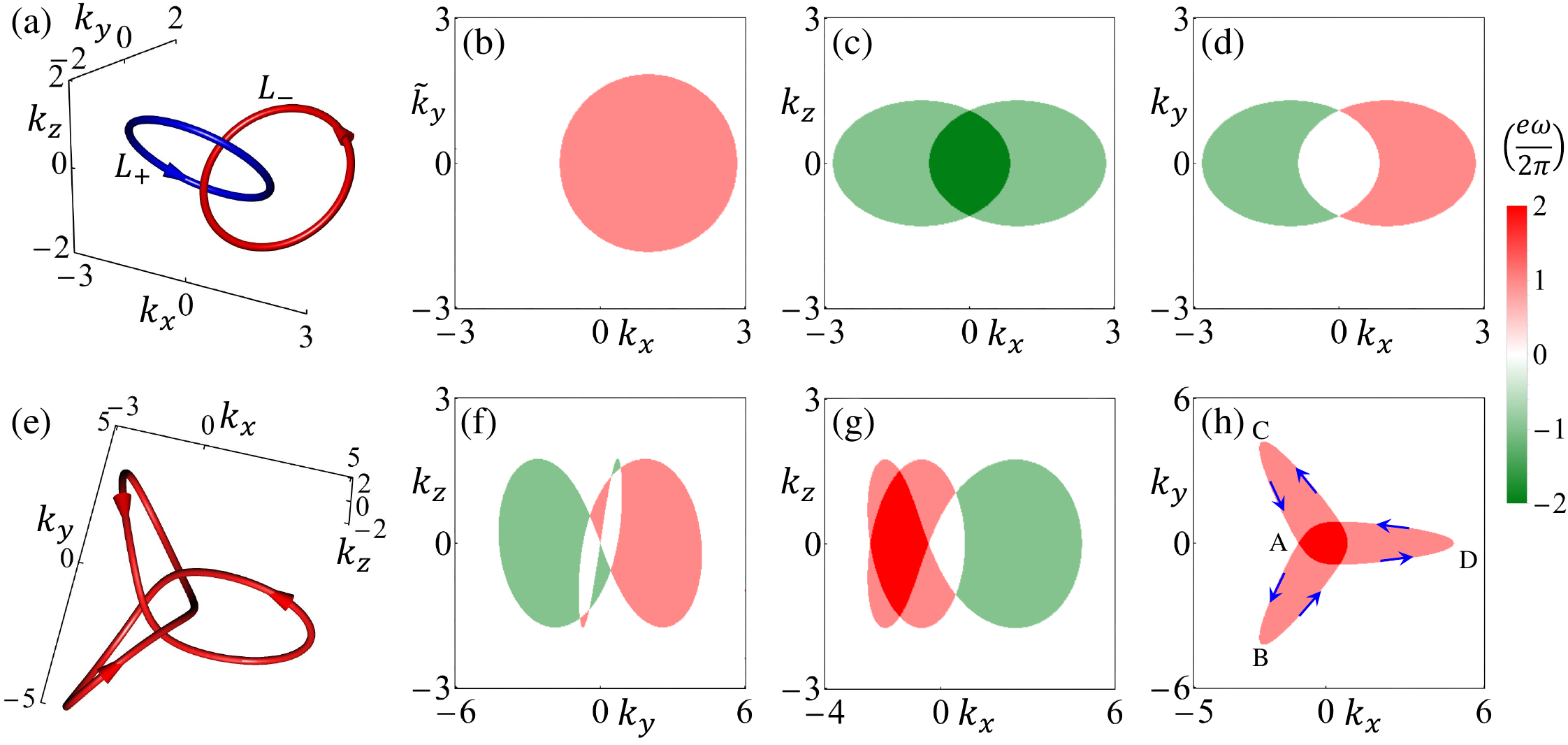}\\[0pt]
\caption{(a) Berry curvature with a Hopf link structure. The distribution of the
pumping current density in (b) $\hat{\bm{n}}_+$, (c) $y$, and (d) $z$ directions
in the corresponding projected reciprocal spaces.
(e) Berry curvature with a knot structure. The distribution of the pumping current density
in (f) $x$, (g) $y$, and (h) $z$ directions. The
parameters are set to $g_1=g_2=1$, and (a-d) $m_0=0.2$, (e-h) $m_0=0.1$.
}
\label{Fig.3}
\end{center}
\end{figure}

It is straightforward to extend the EMI effect to more complex configurations
of the solenoidal Berry curvature, such as links and knots. These novel field configurations have been
explored in various contexts, including electromagnetic fields~\cite{ranada1989topological,
ranada1990knotted,kamchatnov1982topological, ranada1996ball, faddeev1997stable,Battye98prl,radu2008stationary,Kawaguchi08prl,hall2016tying}.
Correspondingly, similar objects associated with the Berry curvature in reciprocal space have also been
proposed~\cite{chen17prb,yan17prb,bi17prb}. To investigate the EMI caused by the solenoidal
Berry curvature forming a Hopf link, we employ the following Hamiltonian
\begin{equation}\label{ham link}
\begin{split}
H_{\text{link}}(\bm{k},t)=&\left[2k_xk_z+2k_ym(t)\right]\sigma_x+g_y(t)\sigma_y\\
&+\left[k_x^2+k_y^2-k_z^2-m^2(t)\right]\sigma_z,
\end{split}
\end{equation}
where we set all coefficients to unity for simplicity, and $m(t)=m_0+g_z(t)-k^2/2$.
We assume that $g_2>m_0+1/2$ to ensure that the conduction and
valence bands periodically intersect at $t=2n\pi/\omega$ along two linked nodal loops.
These loops are defined by parametric equations: $k_y=k_z$,
$(k_x-1)^2+k_y^2+k_z^2=2(m_0+g_2)+1$ for one loop, and $k_y=-k_z$,
$(k_x+1)^2+k_y^2+k_z^2=2(m_0+g_2)+1$ for the other;
see Fig.~\ref{Fig.3}(a).

Similar to the situation with a single loop, the Berry curvature exhibits divergence along the linked nodal loops
at $t=2n\pi/\omega$. The occurrence of this singularity is accompanied by a $2\pi$ jump in
the Berry flux, which induces an $\bm{\mathcal{E}}$ field for each loop in the respective direction.
Specifically, the two nodal loops $L_\pm$, situated in the planes $k_y=\pm k_z$ as depicted Fig.~\ref{Fig.3}(a), generate
average pumping currents in the directions $\hat{\bm{n}}_\pm=(0,-1,\pm 1)$, which are perpendicular to the
corresponding planes. The total current is a joint contribution from both nodal loops. We
consider a fully occupied valence band
and calculate the pumping current in the $\hat{\bm{n}}_+$ direction. In this case,
only the $L_+$ loop, with its plane perpendicular to $\hat{\bm{n}}_+$, contributes to the current.
It is insightful to examine the current distribution in each channel labelled by
$(k_x,\tilde{k}_y)$, where $\tilde{k}_y=(k_y+k_z)/\sqrt{2}$; see Fig.~\ref{Fig.3}(b).
One can see that each 1D channel within the nodal loop contributes a current density of $e\omega/(2\pi)$.
Consequently, the total current density is given by $\bm{J}=e\omega S/(8\pi^3)\hat{\bm{n}}_+$
with $S=\pi[2(m_0+g_z)+1]$ representing the area encircled by $L_+$, consistent with
the findings for a single nodal loop. Similarly, one can derive the current distribution in the $\hat{\bm{n}}_-$
direction, which is solely induced by the nodal loop $L_-$.

In a general direction, the pumping current contains the contribution from both nodal loops, $L_\pm$.
The channel-resolved current density distributions in the $y$ and $z$ directions are shown in
Figs.~\ref{Fig.3}(c,d).
Within the projection area encircled by the linked nodal loops, the current density
is quantized for each channel. In Figs.~\ref{Fig.3}(c,d), there are
two types of regions where the projections
of the two loops either overlap or do not.
In the non-overlapping region, the current density is $\pm e\omega/(2\pi)$ for each channel
with its sign determined by the direction of the $\bm{\mathcal{E}}$ field induced by
the corresponding nodal loop. In the overlapping region, the pumping
current in each channel consists of the contributions from both loops:
it doubles due to the
same direction of the currents induced by the two loops in Fig.~\ref{Fig.3}(c),
whereas the currents induced by the two nodal
loops cancel each other out in Fig.~\ref{Fig.3}(d).
The total pumping current is obtained by integrating the contributions from
all channels. In the $y$ direction, the total current density is $J_y=-e\omega S_1/(4\pi^3)$,
where $S_1=S/\sqrt{2}$ represents the projected area of one nodal loop.
By contrast, the current density in the $z$ direction is zero because
the currents induced by the two loops are opposite in sign. Moreover, the current
in the $x$ direction also vanishes simply because the projection
of both loops does not enclose any area.

The EMI by the varying Berry curvature forming a knot structure can
be described by the following model
\begin{equation}\label{ham knot}
\begin{split}
H_{\mathrm{knot}}(\bm{k},t)&=[k_x^3-3k_xk_y^2+k_z^2-m^2(t)]\sigma_x\\
&+g_y(t)\sigma_y+[3k_x^2k_y-k_y^3+2k_zm(t)]\sigma_z.
\end{split}
\end{equation}
Again, the SGT plays a central role in the topological pumping effect.
At $t=2n\pi/\omega$, band degeneracy occurs periodically along the nodal loop that forms a knot
as shown in Fig.~\ref{Fig.3}(e). This is accompanied by a $2\pi$ jump in the Berry flux.
We calculate the distribution of the pumping current density along the $x$, $y$,
and $z$ directions; see Figs.~\ref{Fig.3}(f-h). One can see that the regions
exhibiting nontrivial current distributions are encircled by the projection of the nodal knot.
The rich structure of the current distributions reveals the complex configurations of a knot.
The pumping current associated with a specific momentum channel
is determined by multiplying $e\omega/(2\pi)$ by the winding number,
which is calculated by tracing the projection of the arrowed nodal knot
in the plane surrounding the reference momentum point. Using Fig.~\ref{Fig.3}(h)
as an example, the projected nodal knot follows the path denoted by the
blue arrows, $ABACADA$.
In the central region, the winding number
is two, while in the three petal-shaped regions, it equals one. It can be
confirmed that the distribution of pumping current in different directions
follows the same rule when meticulously tracking the projected trajectory
of the nodal knot. Moreover, the interpretation of pumping current through the
winding number holds true for all examples presented in this study. In the
case of the nodal link, the total winding number is the sum of those
defined by the two individual loops.

The anomalous EMI effect can be hopefully implemented in
nodal-line semimetals~\cite{Kim15prl,Yu15prl,Heikkila11jetp,
Weng15prb,Chen15nl,Zeng15arxiv,Fang15prb,
Yamakage16jpsj,Xie15aplm,Chan16prb,Zhao16prb,
Bian16prb,Bian16nc} that possess both spin-orbit coupling and magnetic order.
In such a scenario, the Pauli matrices in the Hamiltonian~\eqref{Ham NLSM} represent the
real electronic spin. The  nodal loop naturally carries
a singular Berry flux of $\pm\pi$, which is the central ingredient for the anomalous EMI.
Moreover, the driving terms $g_{y,z}(t)$ can be achieved by
the precessing magnetization controlled by ferromagnetic resonance~\cite{Kittel48pr},
a well-established technique in spin pumping~\cite{Tserkovnyak05rmp,ando2011inverse}.
The EMI can then be probed by the pumping current driven by the microwave.
Interestingly, recent advancements have led to the
observation of nodal link semimetals in a magnetic material~\cite{belopolski2022observation,Chang17prl},
which opens the possibility for the implementation of
anomalous EMI effect induced by solenoidal Berry curvature with link and knot configurations.
Additionally, it is also of particular interest to explore such exotic electromagnetic phenomena
in engineered systems such as ultracold atoms~\cite{lohse2018exploring} and photonics~\cite{zilberberg2018photonic}.

\begin{acknowledgments}
This work was supported by the National Natural Science Foundation of
China under Grant No.\ 12074172 (W.C.), No.\ 12222406 (W.C.),
and No.\ 12264019 (W.L.), and the State Key Program for Basic
Researches of China under Grants No. 2017YFA0303203 (D.Y.X.).
\end{acknowledgments}


%

\end{document}